\documentclass[doublespacing]{elsart}

\usepackage{amsmath}
\usepackage{graphicx}
\usepackage{algorithm}
\usepackage{algorithmic}

\begin{document}

\bibliographystyle{elsart-num}

\newcommand{\ord}[1]{\mathcal{O}(#1)}
\newcommand{\rl}[1]{#1_{12}}
\newcommand{\eps}{\epsilon}
\newcommand{\rhat}{\hat{\mathbf{r}}_{12}}
\newcommand{\nhat}{\hat{\mathbf{n}}}
\newcommand{\rhohat}{\hat{\mathbf{\rho}}}
\newcommand{\dr}{\delta_{R}}
\newcommand{\dt}{\delta_{T}}
\newcommand{\resq}{RE-squared~}

\begin{frontmatter}

\title{Analytical First Derivatives of the \resq Interaction Potential}

\address[sharif]{Sharif University of
Technology, Department of Physics, P.O. Box 11365-9161, Tehran,
Iran.}
\address[max]{Max-Planck-Institut f\"ur Physik komplexer
Systeme, N\"othnitzer Str. 38, 01187 Dresden, Germany}

\author[sharif]{M. Babadi}
\author[sharif]{M. R. Ejtehadi}
\ead{ejtehadi@sharif.edu}
\author[max]{R. Everaers}

\begin{abstract}
We derive exact expressions for the forces and torques between
biaxial molecules interacting via the \resq potential, a recent
variant of the Gay-Berne potential. Moreover, efficient routines
have been provided for rigid body MD simulations, resulting in 1.6
times speedup compared to the two-point finite difference approach.
It has also been shown that the time cost of a MD simulation will be
almost equal to a similar MC simulation, making use of the provided
routines.
\end{abstract}


\end{frontmatter}

\section{Introduction}
In molecular simulations, the van der Waals interactions have a
prominent and essential contribution to the non-bonded interactions
and are typically described using the Lennard-Jones(6-12) potential
or its variants~\cite{AllenTildesley,Frenkel}. An interaction
potential of this type between two extended molecules is assumed to
be a double summation over the respective atomic interaction sites:
\begin{equation}\label{eq:UMacroSum}
U_{int}(\mathcal{M}_1, \mathcal{M}_2) =
\sum_{i\in\mathcal{M}_1}\sum_{j\in\mathcal{M}_2} U_a(r_{ij}; i, j)
\end{equation}
where $\mathcal{M}_1$ and $\mathcal{M}_2$ denote the interacting
molecules and $U_a(\cdot)$ is the atomic interaction potential, e.g
the Lennard-Jones(6-12) potential. The required computation time for
the exact evaluation of this double sum is quadratic in the number
of interacting sites. In practice, a large distant interaction
cutoff accompanied by a proper tapering is used to reduce the
computation cost. More sophisticated and efficient approximate
summation methods such as Ewald summation and the Method of
Lights~\cite{Leach} are also widely used.

As an alternative approach, Gay and Berne~\cite{GB} proposed a
more complicated single-site interaction potential (in contrast to
a more sophisticated summation) for uniaxial rigid molecules which
was generalized to dissimilar and biaxial molecules by Berardi
{\it et al} as well~\cite{BFZ98}. In response to the criticism of
the unclear microscopic interpretation of the Gay-Berne
potential~\cite{Perram96}, we have recently used results from
colloid science to derive an interaction potential through a
systematic approximation of the Hamaker integral~\cite{Hunter} for
mixtures of ellipsoids of arbitrary size and shape, namely the
\resq potential. The parameter space of the \resq potential is
almost identical to that of Berardi, Fava and
Zannoni~\cite{BFZ98}, agrees significantly better with the
numerically evaluated continuum approximation of
Eq.~(\ref{eq:UMacroSum}) and has no nonphysical large distant
limit. It has been verified that the new potential is superior to
the biaxial Gay-Berne potential in representing the atomistic
interactions of small organic molecules as well~\cite{Babadi2006}.
Moreover, the potential of mean force is representable with the
same functional form of the \resq potential with negligible
error~\cite{Babadi2006}.

In an anisotropic coarse-grained potential model, a molecule
($\mathcal{M}$) is treated and described like a rigid body, leading
to a considerable speedup in numerical simulations while preserving
the fundamental behavior of atomisic potentials. Neglecting the
atomic details, each molecule is characterized by the position of
its center (a vector $\mathbf{r}$) and its orientation (a unitary
operator~$\mathbf{A}$ or a unit quaternion~$q$).

Due to the complexity of the functional form of such potentials,
numerical finite differences are widely used for the evaluation of
forces and torques in rigid~body molecular dynamics simulations. The
numerical differentiation methods are prone to round-off errors and
are generally expensive in large scale simulations.

In this article, we will derive analytic expressions for the forces
and torques between two molecules interacting via the \resq
potentials. A set of optimized routines will be suggested for an
efficient implementation of the given expressions. Finally, a time
comparison between the two-point finite difference and the analytic
derivatives will be presented.

\section{The \resq Potential}
As mentioned earlier, the \resq potential~\cite{EE03} is a
coarse-grained description of the attractive and repulsive
interactions between two biaxial molecules. Each molecule is
treated like a biaxial ellipsoid and is described by two
characteristic diagonal tensors (in the principal basis of the
molecule) $\mathbf{S}$ and~$\mathbf{E}$, representing the half
radii of the molecule and the strength of the pole contact
interactions, respectively. As mentioned earlier, the orientation
of a molecule is described by a center displacement vector
$\mathbf{r}$ and a unitary operator~$\mathbf{A}$, revolving the
bases of lab frame to the principal frame of the molecule.

The attractive and repulsive contributions of the \resq potential
between two molecules with a relative center displacement of
$\rl{\mathbf{r}}= \mathbf{r}_2 - \mathbf{r}_1$ and respective
orientation tensors $\mathbf{A}_1$ and $\mathbf{A}_2$ are
respectively:
\begin{subequations}\label{eq:re2}
\begin{multline}\label{eq:re2A}
U_A^{\rm RE-squared}(\mathbf{A}_1, \mathbf{A}_2, \rl{\mathbf{r}})
=-\frac{A_{12}}{36}\Big(1+
3\rl{\eta}\rl{\chi}\frac{\sigma_c}{\rl{h}}\Big)\times\\
\prod_{i=1}^2\prod_{e=x,y,z}
\Bigg(\frac{\sigma^{(i)}_e}{\sigma^{(i)}_e+h_{12}/2}\Bigg)
\end{multline}
\begin{multline}\label{eq:re2R}
U_R^{\rm RE-squared}(\mathbf{A}_1, \mathbf{A}_2,
\rl{\mathbf{r}})=\frac{A_{12}}{2025}
\Big(\frac{\sigma_c}{\rl{h}}\Big)^6\Big(1+
\frac{45}{56}\rl{\eta}\rl{\chi}\frac{\sigma_c}{\rl{h}}\Big)\times\\
\prod_{i=1}^2 \prod_{e=x,y,z}
\Bigg(\frac{\sigma^{(i)}_e}{\sigma^{(i)}_e+h_{12}/60^{\frac{1}{3}}}\Bigg)
\end{multline}
\end{subequations}
where $A_{12}$ is the Hamaker constant (the energy scale),
$\sigma_c$ is the atomic interaction radius and $\sigma_x^{(i)}$,
$\sigma_y^{(i)}$ and $\sigma_z^{(i)}$ are the half-radii of $i$th
ellipsoid (i=1,2). $\rl{\eta}$ and $\rl{\chi}$ are purely
orientation dependant terms, describing the anisotropy of the
molecules and $\rl{h}$ is the the least contact distance between the
ellipsoids.

The structure tensor $\mathbf{S}_i$ and the relative well-depth
tensor $\mathbf{E}_i$ are diagonal in the principal basis of $i$th
molecule and are defined as:
\begin{subequations}
\begin{equation}
\mathbf{S}_i = \textrm{diag}\{\sigma_x^{(i)}, \sigma_y^{(i)},
\sigma_z^{(i)}\}
\end{equation}
\begin{equation}
\mathbf{E}_i = \textrm{diag}\left\{E_x^{(i)}, E_y^{(i)},
E_z^{(i)}\right\}
\end{equation}
\end{subequations}
where $E_x^{(i)}$, $E_y^{(i)}$ and $E_z^{(i)}$ are dimensionless
energy scales inversely proportional to the well-depths of the
respective orthogonal configurations of the interacting molecules.
For large molecules with uniform constructions, it has been
shown~\cite{EE03} that the energy parameteres are approximately
representable in terms of the local contact curvatures using the
Derjaguin expansion~\cite{EE03,Derjaguin}:
\begin{equation}\label{eq:derya}
\mathbf{E}_i = \sigma_c \textrm{diag}\left\{\frac{\sigma_x}{\sigma_y
\sigma_z}, \frac{\sigma_y}{\sigma_x \sigma_z},
\frac{\sigma_z}{\sigma_x \sigma_y}\right\}
\end{equation}
The term $\rl{\chi}$ quantifies the strength of interaction with
respect to the local atomic interaction strength of the molecules
and is defined as:
\begin{equation}\label{eq:chidef}
\rl{\chi}(\mathbf{A}_1, \mathbf{A}_2, \rhat) =
2\rhat^T\rl{\mathbf{B}}^{-1}(\mathbf{A}_1, \mathbf{A}_2)\rhat
\end{equation}
where $\rl{\mathbf{B}}$ is defined in terms of the orientation
tensors $\mathbf{A}_i$ and relative well-depth tensors
$\mathbf{E}_i$:
\begin{equation}\label{eq:Bdef}
\rl{\mathbf{B}}(\mathbf{A}_1, \mathbf{A}_1) = \mathbf{A}_1^T
\mathbf{E}_1 \mathbf{A}_1 + \mathbf{A}_2^T \mathbf{E}_2
\mathbf{A}_2.
\end{equation}
The term $\rl{\eta}$ describes the effect of contact curvatures of
the molecules in the strength of the interaction and is defined as:
\begin{equation}\label{eq:etadef}
\rl{\eta}(\mathbf{A}_1, \mathbf{A}_2, \rhat) =
\frac{\det[\mathbf{S}_1]/\sigma_1^2 +
\det[\mathbf{S}_2]/\sigma_2^2}{\big[\det[\rl{\mathbf{H}}]/(\sigma_1+\sigma_2)\big]^{1/2}},
\end{equation}
Where $\sigma_i$ is the projected radius of \textit{i}th ellipsoid
along $\rhat$:
\begin{equation}\label{eq:sigdef}
\sigma_i(\mathbf{A}_i,
\rhat)=(\rhat^T\mathbf{A}_i^T\mathbf{S}_i^{-2}\mathbf{A}_i\rhat)^{-1/2}
\end{equation}
and the tensor $\rl{\mathbf{H}}$ is defined as:
\begin{equation}\label{eq:Hdef}
\rl{\mathbf{H}}(\mathbf{A}_1,\mathbf{A}_2,\rhat) =
\frac{1}{\sigma_1} \mathbf{A}_1^T \mathbf{S}_1^2 \mathbf{A}_1 +
\frac{1}{\sigma_2} \mathbf{A}_2^T \mathbf{S}_2^2 \mathbf{A}_2 .
\end{equation}
No trivial solution is available for the least contact distance
between two arbitrary ellipsoids ($\rl{h}$)~\cite{Perram96}. The
Gay-Berne approximation~\cite{GB,Perram96} is usually employed due
to its low complexity and acceptable performance:
\begin{equation}\label{eq:hgb}
\rl{h}^{GB} = \|\rl{\mathbf{r}}\| - \rl{\sigma},
\end{equation}
where the anisotropic distance function $\rl{\sigma}$~\cite{BFZ98}
is defined as:
\begin{equation}\label{eq:sig12def}
\rl{\sigma}=\left(\frac{1}{2}\rhat^T \rl{\mathbf{G}}^{-1}
\rhat\right)^{-\frac{1}{2}}
\end{equation}
and the symmetric overlap tensor $\rl{\mathbf{G}}$ is:
\begin{equation}\label{eq:Gdef}
\rl{\mathbf{G}} = \mathbf{A}_1^T\mathbf{S}_1^2\mathbf{A}_1 +
\mathbf{A}_2^T\mathbf{S}_2^2\mathbf{A}_2
\end{equation}
We will also employ this approximation in our derivation and will
omit the superscript GB for shorthand in the rest of the article.

\section{Analytic expressions for forces and torques}
The algebraic structure of the attractive and repulsive
contributions of the \resq potential are essentially the same. Thus,
both of the contributions are expressible with a proper template
structure, defined as:
\begin{equation}\label{eq:generalresq}
U_{\alpha}^{\rm RE-squared}=\frac{\rl{A}}{A_{\alpha}}\Big(\frac{\sigma_c}{\rl{h}}\Big)^{n_{\alpha}}
\Big(1+b_{\alpha}\rl{\eta}\rl{\chi}\frac{\sigma_c}{\rl{h}}\Big)\times
\prod_{i=1}^2
\prod_{e=x,y,z}\Bigg(\frac{\sigma^{(i)}_e}{\sigma^{(i)}_e+h_{12}/c_{\alpha}}\Bigg)
\end{equation}
We will work through this template in the derivation. One may yield
to the explicit form of each of the contributions by giving
appropriate values to the $\alpha$-superscripted parameters
according to Eq.~(\ref{eq:re2A}) and~(\ref{eq:re2R}).

In an interaction between the molecules $\mathcal{M}_1$ and
$\mathcal{M}_2$, the exerted force and torque on the molecule
$\mathcal{M}_2$ is most easily evaluated by applying proper virtual
displacements and infinitesimal rotations to the interaction
potential. The exerted force and torque on $\mathcal{M}_1$ is
trivially obtained using the third law of Newton, afterwards.

We denote the first-order translational and rotational variation
operators on the coordinates of $\mathcal{M}_2$ by $\delta_T$ and
$\delta_R$, respectively. The translational variation operator is
formally defined on a scalar function $F$ as:
\begin{equation}\label{eq:deltaTdef}
\delta_T[F(\mathbf{A}_1, \mathbf{A}_2, \rl{\mathbf{r}}); \rhohat,
\eps] := \eps \frac{\partial}{\partial \eps} F(\mathbf{A}_1,
\mathbf{A}_2, \rl{\mathbf{r}} + \eps \rhohat)\Big|_{\eps=0}
\end{equation}
where the unit vector $\rhohat$ points to the direction of
translational variation and $\epsilon$ is an (infinitesimal) scalar
for which the variations are valued. The proper definition of the
rotational variation is more involved. The projection of the exerted
torque on $\mathcal{M}_2$ along the unit vector $\nhat$ is obtained
by applying the infinitesimal orthogonal operator~
$\mathbf{I}+\eps\hat{\mathbf{n}}.\mathbf{\sigma}$ on the operator
revolving {\it the molecule} from the lab frame to its current
frame, i.e. $\mathbf{A}_2^T$. The resulting orientation operator
would be
$\left(\left(\mathbf{I}+\eps\hat{\mathbf{n}}.\mathbf{\sigma}\right)
\mathbf{A}_2^T\right)^T=\mathbf{A}_2-\eps\mathbf{A}_2\hat{\mathbf{n}}.\mathbf{\sigma}$,
according to the anti-symmetry of the principal rotation generators
($\mathbf{\sigma}$). The discussion suggests the definition:
\begin{multline}
\delta_R[F(\mathbf{A}_1, \mathbf{A}_2, \rl{\mathbf{r}}); \nhat, \eps] := \eps \times\\
\frac{\partial}{\partial \eps} F\big(\mathbf{A}_1,
\mathbf{A}_2-\eps\mathbf{A}_2\mathbf{\Omega},\rl{\mathbf{r}}\big)\Big|_{\eps=0}
\end{multline}
where $\Omega=\nhat.\mathbf{\sigma}$ is the rotation generator
corresponding to the direction $\hat{\mathbf{n}}$. Acting
exclusively on the coordinates of the second molecule
($\mathcal{M}_2$), the operators may be used to define the exerted
force and torque along $\rhohat$ and around $\nhat$ respectively as:
\begin{subequations}\label{eq:FNdef}
\begin{equation}
\mathbf{F}_{\mathcal{M}_2,\rhohat}=-\frac{\dt
(U_A+U_R)}{\eps}\rhohat
\end{equation}
\begin{equation}
\mathbf{N}_{\mathcal{M}_2,\nhat}=-\frac{\dr (U_A+U_R)}{\eps}\nhat
\end{equation}
\end{subequations}
Applying either of the operators to the template potential
$U_\alpha$, we get:
\begin{multline}\label{eq:dU}
\delta U_{\alpha}/U_{\alpha}= \sigma
b_{\alpha}\frac{\rl{\eta}\delta\rl{\chi}+\delta\rl{\eta}\rl{\chi}}
{\rl{h}+\sigma b_{\alpha}\rl{\chi}\rl{\eta}}-\delta\rl{h} \Big(\frac{n_{\alpha}+1}{\rl{h}}-\\
\frac{1}{\rl{h}+b_{\alpha}\rl{\eta}\rl{\chi}\sigma} +\sum_{i=1}^2
\sum_{e=x,y,z}\frac{1}{c_{\alpha}\sigma^{(i)}_e+\rl{h}}\Big)
\end{multline}
We will complete the derivation by providing explicit expressions
for the first-order variations appearing in Eq.~(\ref{eq:dU}).

\subsection{Derivation of the first-order variations}

\subsubsection{Rotational variation of $\rl{\eta}$}
Applying $\dr$ operator to $\rl{\eta}$ (Eq.~\ref{eq:etadef}) and
dropping off the constant terms, we arrive at:
\begin{eqnarray}\label{eq:dreta}
\frac{\dr\rl{\eta}}{\rl{\eta}}&=&
\frac{1}{2}\frac{\dr\sigma_2}{\sigma_1+\sigma_2}-\frac{1}{2}
\frac{\dr\det{\rl{\mathbf{H}}}}{\det{\rl{\mathbf{H}}}}-
\frac{\dr\sigma_2}{\sigma_2^3}\frac{2\det{\mathbf{S}_2}}
{\det{\mathbf{S}_1}/\sigma_1^2+\det{\mathbf{S_2}}/\sigma_2^2}
\end{eqnarray}
The rotational variation of $\sigma_2$ is:
\begin{eqnarray}\label{eq:drsigma2}
\frac{\dr\sigma_2}{\sigma_2} & = &-\frac{1}{2}\sigma_2^2\dr\left(
\rhat^T\mathbf{A}_2^T\mathbf{S}_2^{-2}\mathbf{A}_2\rhat\right)\\
& = &
\frac{1}{2}\eps\sigma_2^2\rhat^T\big((\mathbf{\mathbf{A}_2\Omega})^T
\mathbf{S}_2^{-2}\mathbf{A}_2+\mathbf{A}_2^T\mathbf{S}_2^{-2}
\mathbf{A}_2\mathbf{\Omega}\big)\rhat\nonumber\\
& = & \eps\sigma_2^2\rhat^T(\mathbf{A}_2\mathbf{\Omega})^T
\mathbf{S}_2^{-2}\mathbf{A}_2\rhat
\end{eqnarray}
We have used the symmetry and the anti-symmetry properties of
$\mathbf{S}^{-2}$ and $\mathbf{\Omega}$, respectively. The
rotational variation of $\dr\rl{\mathbf{H}}$ is required prior to
$\dr\det\rl{\mathbf{H}}$:
\begin{eqnarray}\label{eq:drH}
\dr\rl{\mathbf{H}}&=&-\frac{\eps}{\sigma_2}\big(\mathbf{A}_2^T
\mathbf{S}_2^2 \mathbf{A}_2\mathbf{\Omega}
+(\mathbf{A}_2\mathbf{\Omega})^T
\mathbf{S}_2^2\mathbf{A}_2\big)-\nonumber\\
&&\frac{\dr\sigma_2}{\sigma_2^2}\mathbf{A}_2^T\mathbf{S}_2^2
\mathbf{A}_2.
\end{eqnarray}
It can be easily verified that,
\begin{equation}\label{eq:ddet}
\det[\mathbf{\Gamma} + \eps\mathbf{\Lambda}]=
\det[\mathbf{\Gamma}] + \eps\sum_{i=1}^{\dim\mathbf{\Gamma}}
\det[\mathbf{\Gamma}^{(i)}] + \ord{\eps^2}
\end{equation}
for arbitrary $\mathbf{\Gamma}$ and $\mathbf{\Lambda}$, where
$\mathbf{\Gamma}^{(i)}$ is defined as:
\begin{equation}\label{eq:gamma_i_def}
\mathbf{\Gamma}^{(i)}_{kl} = \left\{
\begin{tabular}{ll}
$\mathbf{\Gamma}_{kl}$ & \quad$k\neq i$\\
$\mathbf{\Lambda}_{il}$ & \quad$k=i$
\end{tabular} \right.
\end{equation}
An explicit expression for $\dr\det\rl{\mathbf{H}}$ is feasible
using Eq.~(\ref{eq:drH}) and Eq.~(\ref{eq:ddet}). Equating
$\mathbf{\Gamma}$ and $\mathbf{\Lambda}$ to $\rl{\mathbf{H}}$ and
$\dr\rl{\mathbf{H}}/\eps$ respectively, the first-order terms of
Eq.~(\ref{eq:ddet}) will evidently be equal to the variation we are
looking for.

\subsubsection{Rotational variation of $\rl{\chi}$}
Using Eq.~(\ref{eq:chidef}), it is straightforward to show that:
\begin{equation}\label{eq:drchi_0}
\frac{\dr\rl{\chi}}{\rl{\chi}} =
\frac{\rhat^T\dr(\rl{\mathbf{B}^{-1}})\rhat}
{\rhat^T\rl{\mathbf{B}^{-1}}\rhat}.
\end{equation}
where:
\begin{equation}\label{eq:drB}
\dr\rl{\mathbf{B}} = -\eps\big[(\mathbf{A}_2\mathbf{\Omega})^T
\mathbf{E}_2 \mathbf{A}_2 + \mathbf{A}_2^T \mathbf{E}_2
\mathbf{A}_2\mathbf{\Omega}\big].
\end{equation}
Using the mathematical relation:
\begin{equation}\label{eq:deltainv}
(\mathbf{\Gamma}+\eps\mathbf{\Lambda})^{-1}=
-\eps\mathbf{\Gamma}^{-1}\mathbf{\Lambda}\mathbf{\Gamma}^{-1} +
\ord{\eps^2}
\end{equation}
for infinitesimal $\epsilon$, together with Eq.~(\ref{eq:drchi_0})
and (\ref{eq:drB}) we finally reach to:
\begin{eqnarray}\label{eq:drchi}
\dr\rl{\chi}&=& 2 \eps \rhat^T\rl{\mathbf{B}^{-1}}\big[
(\mathbf{A}_2\mathbf{\Omega})^T\mathbf{E}_2 \mathbf{A}_2+\nonumber\\
&&\mathbf{A}_2^T\mathbf{E}_2\mathbf{\mathbf{A}_2\Omega}\big]\rl{\mathbf{B}^{-1}}\rhat\nonumber\\
&=&4 \eps\rhat^T\rl{\mathbf{B}^{-1}}\mathbf{A}_2^T
\mathbf{E}_2\mathbf{\mathbf{A}_2\Omega}\rl{\mathbf{B}^{-1}}\rhat
\end{eqnarray}
where the symmetry of $\mathbf{E}_i$, $\rl{\mathbf{B}}$ and their
inverses have been used.

\subsubsection{Rotational variation of $\rl{h}$}
We will use the Gay-Berne approximation for the least constant
distance defined by Eq.~(\ref{eq:hgb}). Accordingly, the rotational
variation of $\rl{h}$ is a result of the change in the anisotropic
distance function $\rl{\sigma}$:
\begin{equation}
\dr\rl{h}=\dr{(\rl{r}-\rl{\sigma})}=-\dr\rl{\sigma}
\end{equation}
where the rotationally constant term $\rl{r}$ drops out. The term
$\dr{\rl{\sigma}}$ is easily expressed in terms of
$\dr\rl{\mathbf{G}^{-1}}$:
\begin{equation}\label{eq:drh}
\frac{\dr{\rl{\sigma}}}{{\rl{\sigma}}}=-\frac{1}{2}\frac{\dr{\rhat^T
\rl{\mathbf{G}^{-1}}\rhat}}{\rhat^T\rl{\mathbf{G}^{-1}}\rhat}=
-\frac{1}{4}\rl{\sigma}^2\rhat^T(\dr\rl{\mathbf{G}^{-1}})\rhat
\end{equation}
Eq.~(\ref{eq:Gdef}) together with Eq.~(\ref{eq:deltainv}) result in:
\begin{eqnarray}\label{eq:drGinv}
\dr\rl{\mathbf{G}^{-1}}&=&\eps\rl{\mathbf{G}^{-1}}\big[(\mathbf{A}_2
\mathbf{\Omega})^T \mathbf{S}_2^2 \mathbf{A}_2+
\mathbf{A}_2^T\mathbf{S}_2^2\mathbf{\mathbf{A}_2\Omega}\big]\rl{\mathbf{G}^{-1}}\nonumber\\
&=&2\eps\rl{\mathbf{G}^{-1}}\big[\mathbf{A}_2^T\mathbf{S}_2^2\mathbf{A}_2\mathbf{\Omega}
\big]\rl{\mathbf{G}^{-1}}
\end{eqnarray}
where the symmetry of $\mathbf{S}_2$, $\rl{\mathbf{G}}$ and their
inverses have been used. Thus, we finally reach to:
\begin{equation}\label{eq:drhfinal}
\dr{\rl{h}}=\frac{1}{2}\eps\rl{\sigma}^3\rhat^T\rl{\mathbf{G}^{-1}}\mathbf{A}_2^T
\mathbf{S}_2^2\mathbf{\mathbf{A}_2\Omega}\rl{\mathbf{G}^{-1}}\rhat
\end{equation}

\subsubsection{Translational variation of $\rl{\eta}$}
The displacement of the molecule $\mathcal{M}_2$ results in a change
in the direction of the connecting vector $\rl{\mathbf{r}}$. Up to
first order, this change is expressed as:
\begin{eqnarray}\label{eq:deltarhat_0}
\rhat^{(1)}&=&\frac{\rl{\mathbf{r}}^{(0)}+\eps\rhohat}
{\|\rl{\mathbf{r}}^{(0)}+\eps\rhohat\|}\nonumber\\
&=&\rhat^{(0)} +
\frac{\eps}{\rl{r}}\big(\rhohat-(\rhohat.\rhat)\rhat\big)+
\ord{\eps^2}
\end{eqnarray}
Defining a new auxiliary vector results in a cleaner derivation:
\begin{equation}\label{eq:beta}
\mathbf{u}:=\frac{\rhohat-(\rhohat.\rhat)\rhat}{\rl{r}}
\end{equation}
Accordingly, $\delta\rhat$ is obviously $\eps$ times $\mathbf{u}$.
Applying $\dt$ operator to $\rl{\eta}$, we reach to:
\begin{eqnarray}\label{eq:dteta}
\frac{\dt\rl{\eta}}{\rl{\eta}}&=&
\frac{1}{2}\frac{\dt\sigma_1+\dt\sigma_2}{\sigma_1+\sigma_2}-\frac{1}{2}
\frac{\dt\det{\rl{\mathbf{H}}}}{\det{\rl{\mathbf{H}}}}-\nonumber\\
&&2\frac{(\dt\sigma_1)\det{\mathbf{S}_1}/\sigma_1^3
+(\dt\sigma_2)\det{\mathbf{S}_2}/\sigma_2^3}
{\det{\mathbf{S}_1}/\sigma_1^2+\det{\mathbf{S_2}}/\sigma_2^2}
\end{eqnarray}
We will follow the same steps as the rotational case. The
translational variation of the projected diameter $\sigma_i$ is:
\begin{eqnarray}\label{eq:dtsigma_i}
\dt\sigma_i&=&-\frac{1}{2}\sigma_i^3\dt\left(\rhat^T\mathbf{A}_i^T\mathbf{S}_i^{-2}
\mathbf{A}_i\rhat\right)\nonumber\\
&=&-\frac{1}{2}\eps\sigma_i^3
\left(\mathbf{u}^T\mathbf{A}_i^T\mathbf{S}_i^{-2}\mathbf{A}_i\rhat+\rhat^T\mathbf{A}_i^T
\mathbf{S}_i^{-2}\mathbf{A}_i\mathbf{u}\right)\nonumber\\
&=&-\eps\sigma_i^3\mathbf{u}^T
\mathbf{A}_i^T\mathbf{S}_i^{-2}\mathbf{A}_i\rhat
\end{eqnarray}
It is also easy to verify that:
\begin{equation}\label{eq:dtH}
\dt\rl{\mathbf{H}}=-\frac{\dt\sigma_1}{\sigma_1^2}\mathbf{A}_1^T
\mathbf{S}_1^2 \mathbf{A}_1
-\frac{\dt\sigma_2}{\sigma_2^2}\mathbf{A}_2^T \mathbf{S}_2^2
\mathbf{A}_2
\end{equation}
Finally, we may express $\dt\det[\rl{\mathbf{H}}]$ explicitly using
Eq.~(\ref{eq:ddet}) in terms of $\rl{\mathbf{H}}$ and its
translational variation, Eq.~(\ref{eq:dtH}).

\subsubsection{Translational variation of $\rl{\chi}$}
Applying $\dt$ to $\rl{\chi}$, we get:
\begin{equation}\label{eq:dtchi}
\frac{\dt\rl{\chi}}{\rl{\chi}}=
\frac{\dt\rhat^T\rl{\mathbf{B}^{-1}}\rhat}
{\rhat^T\rl{\mathbf{B}^{-1}}\rhat}
\end{equation}
The numerator simplifies to:
\begin{eqnarray}
\dt\rhat\rl{\mathbf{B}^{-1}}\rhat&=&
\eps\big(\mathbf{u}^T\rl{\mathbf{B}^{-1}}\rhat+
\rhat^T\rl{\mathbf{B}^{-1}}\mathbf{u}\big)\nonumber\\
&=&2\eps\mathbf{u}^T\rl{\mathbf{B}^{-1}}\rhat
\end{eqnarray}
We finally reach to:
\begin{equation}\label{eq:dtchifinal}
\dt\rl{\chi}=4 \eps\mathbf{u}^T\rl{\mathbf{B}^{-1}}\rhat
\end{equation}
We have used the symmetry and the translational invariance of
$\rl{\mathbf{B}}$.

\subsubsection{Translational variation of $\rl{h}$}
Both of the involving terms in the definition of $\rl{h}$ contribute
to $\dt\rl{h}$. The contribution of the center displacement is:
\begin{equation}\label{eq:dtr}
\dt\rl{r}
=\eps\rhat.\rhohat
\end{equation}
and the variation of the anisotropic distance function may be
expressed as:
\begin{equation}\label{eq:dtsigma12}
\frac{\dt\rl{\sigma}}{\rl{\sigma}}=-\frac{1}{2}
\frac{\dt\rhat^T\rl{\mathbf{G}^{-1}}\rhat}
{\rhat^T\rl{\mathbf{G}^{-1}}\rhat}
\end{equation}
Expanding and simplifying the numerator, we reach to:
\begin{equation}\label{eq:dtsigma12final}
\dt\rl{\sigma}=-\frac{1}{2}\eps\rl{\sigma}^3
\mathbf{u}^T\rl{\mathbf{G}^{-1}}\rhat
\end{equation}
Adding up the above contributions, we finally get:
\begin{equation}
\dt\rl{h}=\eps\rhat.\rhohat+\frac{1}{2}\eps\rl{\sigma}^3
\mathbf{u}^T\rl{\mathbf{G}^{-1}}\rhat
\end{equation}

\section{An Efficient Implementation for Rigid-body Molecular Dynamics Simulations}
\begin{figure}[!h]
\center
\includegraphics[bb=40pt 195pt 560pt 600pt, scale=0.5]{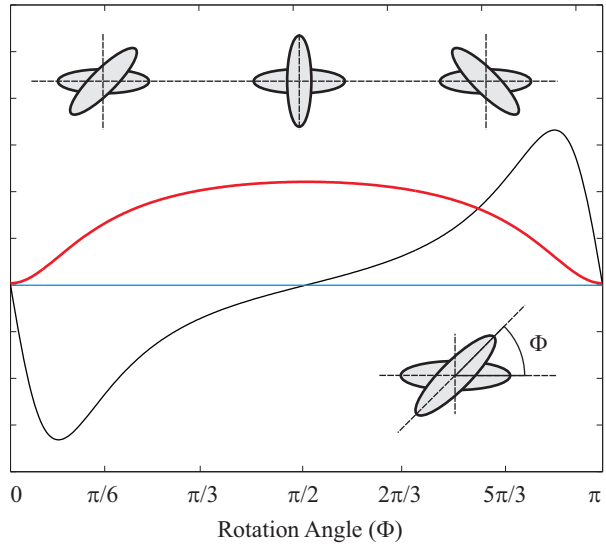}
\vspace{10pt} \caption{Typical interaction energy (red thick line)
and torque (black thin line) between two prolate molecules (in
arbitrary units), continuously evolved from side-by-side to cross
configuration. The ellipsoids are identical, having half-radii [11 :
2 : 0.5] (in arbitrary units) and vertically separated by 5 units of
length.} \label{fig:torque}
\end{figure}

\begin{figure}[!h]
\center
\includegraphics[bb=40pt 195pt 560pt 600pt, scale=0.5]{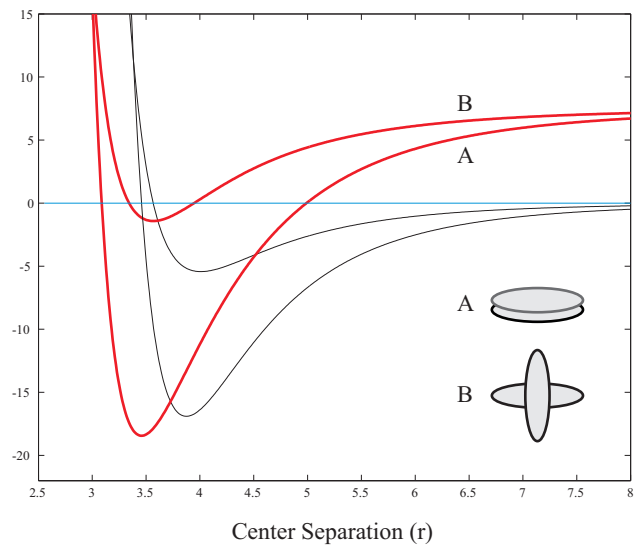}
\vspace{10pt} \caption{Interaction energy (red thick lines) and
vertical force (black thin lines) between two prolate molecules in
two different configurations with respect to the vertical center
separation (r). The ellipsoids are identical, having half-radii [11
: 2 : 0.5] (in arbitrary units). The y-axis ticks correspond to the
vertical force (in arbitrary units).} \label{fig:force}
\end{figure}

Most of the required matrix and vectors products in the evaluation
of the first derivatives using the given expressions will be already
available once one gets through the evaluation of the interaction
energy beforehand. Without a careful implementation, a minimal
speedup is expected due to the considerable redundancy of the
algebra. Therefore, a proper integration between the variable spaces
of all routines must be considered. The three provided routines
demonstrate a suggested implementation. The first routine evaluates
the interaction energy while the second and third routines calculate
the torque and force. The latter routines depend on portions of
variable space of the first routine in order to skip the redundant
matrix products. We have also omitted the $\eps$ factors appearing
in the variations beforehand as they will finally factor out,
according to Eq.~\ref{eq:FNdef}. In practice, one call of the first
routine accompanied by three calls of each of the second and third
routines are mandatory in order to evaluate the three components of
the force and the torque vectors. We have compared the computation
time of an efficient C-language implementation of the proposed
routines~\cite{RE2CLib} against a numerical two-point finite
differentiation. A large scale comparison (Pentium-M 1.7Ghz, GCC4)
indicates that the an evaluation of the interaction energy and the
force and torque vectors takes 38.6~$\mu$s using the provided
routines while the same calculation takes 62.2~$\mu$s with the
finite difference approach, leading to 1.6 times speedup.
Figures~(\ref{fig:torque}) and~(\ref{fig:force}) have been drawn
with the aid of the provided routines and show the typical behavior
of interaction force and torque between two prolate molecules.

\section{Monte-Carlo and Molecular Dynamics Simulations Time Cost}
Monte Carlo (MC) and Molecular Dynamics (MD) simulations are two
main concerns in studying molecular systems. MC simulations are
usually faster and more effective in the studying of steady states
while MD simulations play a more prominent role in the studying of
transition states of certain systems. Furthermore, there are certain
cases where MC simulations are of little interest (specially where
the dynamical behavior is demanded).

A MC step is considered to be as {\it revealing} an a MD step once
each molecule successfully move to a new position in the phase
space. In a system of N molecules, each having an average number of
M neighbors, the time consumption of a MC step roughly is:
\begin{equation}
T_{MC} = \alpha N M \times \tau_E
\end{equation}
where $\alpha$ is the inverse of the acceptance ratio (usually,
$\alpha \simeq 2$ with a proper conditioning) and $\tau_E$ is the
average required time of an energy evaluation. The corresponding
time consumption of an MD step would be:
\begin{equation}
T_{MD} = \frac{NM}{2}(\tau_E + \tau_F + \tau_T)
\end{equation}
where $\tau_F$ and $\tau_T$ are the average excessive time required
for a single force and torque evaluation in all three directions.
Using the values obtained from a sample large-scale simulation (with
an acceptance of 50\%), the ratio of the time expenses turn out to
be:
\begin{equation}\label{MCMD}
\frac{T_{MD}}{T_{MC}} = \frac{38.6 \quad (\mu s)}{2 \times 2 \times
8.9 \quad (\mu s)} \simeq 1.1
\end{equation}
using analytical first derivatives. The same ratio would be 1.7
using finite differences. Therefore, one will end up with a MD
simulation almost as fast as a MC simulation using the provided
analytical derivatives.

\newpage
\begin{center}
Routine 1: Evaluation of $U_{\alpha}$
\end{center}
\begin{algorithmic}[1]
\STATE $\rhat \Leftarrow \rl{\mathbf{r}}/\|\rl{\mathbf{r}}\|$ \STATE
$\mathbf{\Gamma}_1 \Leftarrow
\mathbf{A}_1^T\mathbf{S}_1^2\mathbf{A}_1$ \STATE $\mathbf{\Gamma}_2
\Leftarrow \mathbf{A}_2^T\mathbf{S}_2^2\mathbf{A}_2$ \STATE
$\mathbf{s} \Leftarrow
(\mathbf{\Gamma}_1+\mathbf{\Gamma}_2)^{-1}\rhat$ \STATE $\rl{\sigma}
\Leftarrow 1/\sqrt{\frac{1}{2}\mathbf{s}^T\rhat}$ \STATE
$\mathbf{z}_1 \Leftarrow \mathbf{A}_1\rhat$ \STATE $\mathbf{z}_2
\Leftarrow \mathbf{A}_2\rhat$ \STATE $\mathbf{v}_1 \Leftarrow
\mathbf{S}_1^{-2}\mathbf{z}_1$ \STATE $\mathbf{v}_2 \Leftarrow
\mathbf{S}_2^{-2}\mathbf{z}_2$ \STATE $\sigma_1 \Leftarrow
1/\sqrt{\mathbf{z}_1^T\mathbf{v}_1}$ \STATE $\sigma_2 \Leftarrow
1/\sqrt{\mathbf{z}_2^T\mathbf{v}_2}$ \STATE $\rl{\mathbf{H}}
\Leftarrow \mathbf{\Gamma}_1/\sigma_1 + \mathbf{\Gamma}_2/\sigma_2$
\STATE $d_H \Leftarrow \det\rl{\mathbf{H}}$ \STATE $d_{S1}
\Leftarrow \det\mathbf{S}_1$ \STATE $d_{S2} \Leftarrow
\det\mathbf{S}_2$ \STATE $\lambda \Leftarrow d_{S1}/\sigma_1^2 +
d_{S2}/\sigma_2^2$ \STATE $\nu \Leftarrow
\sqrt{d_H/(\sigma_1+\sigma_2)}$ \STATE $\mathbf{w} \Leftarrow
(\mathbf{A}_1^T\mathbf{E}_1\mathbf{A}_1 +
\mathbf{A}_2^T\mathbf{E}_2\mathbf{A}_2)^{-1}\rhat$
\STATE $\rl{h} \Leftarrow \rl{r} - \rl{\sigma}$ \STATE $\rl{\eta}
\Leftarrow \lambda/\nu$ \STATE $\rl{\chi} \Leftarrow
2\rhat^T\mathbf{w}$ \STATE Evaluate $U_\alpha$ using
Eq.~(\ref{eq:generalresq})
\end{algorithmic}

\newpage
\begin{center}
Routine 2: Evaluation of $\dr U_{\alpha}$
\end{center}
\begin{algorithmic}[1]
\REQUIRE Evaluated variable space of Routine~(1) \STATE
$\mathbf{\Lambda} \Leftarrow -\mathbf{A}_2(\nhat.\mathbf{\sigma})$
\STATE $\mathbf{p} \Leftarrow \Lambda\rhat$ \STATE $\dr\sigma_2
\Leftarrow -\sigma_2^3\mathbf{p}^T\mathbf{v}_2$ \STATE
$\dr\rl{\mathbf{H}} \Leftarrow
(\mathbf{A}_2^T\mathbf{S}_2^2\mathbf{\Lambda}+
\mathbf{\Lambda}^T\mathbf{S}_2^2\mathbf{A}_2)/\sigma_2
-(\dr\sigma_2/\sigma_2^2)\mathbf{\Gamma}_2$ \STATE $\dr d_H
\Leftarrow 0$ \FOR{$i=1$ to $3$} \STATE $\mathbf{J} \Leftarrow
\rl{\mathbf{H}}^{(i)}$ \COMMENT{Defined in the corresponding
section} \STATE $\dr d_H \Leftarrow \dr d_H + \det\mathbf{J}$
\ENDFOR \STATE $\dr\rl{\eta} \Leftarrow
\frac{\rl{\eta}\dr\sigma_2}{2(\sigma_1+\sigma_2)}
-\frac{\rl{\eta}\dr d_H}{2 d_H} - \frac{2
\rl{\eta}d_{S2}\dr\sigma_2}{\lambda\sigma_2^3}$ \STATE $\dr\rl{\chi}
\Leftarrow
-4\mathbf{w}^T\mathbf{A}_2^T\mathbf{E}_2\mathbf{\Lambda}\mathbf{w}$
\STATE $\dr\rl{h} \Leftarrow
-\frac{1}{2}\rl{\sigma}^3\mathbf{s}^T\mathbf{\Phi}\mathbf{s}$ \STATE
Evaluate $\dr U_{\alpha}$ using Eq.~(\ref{eq:dU})
\end{algorithmic}

\newpage
\begin{center}
Routine 3: Evaluation of $\dt U_{\alpha}$
\end{center}
\begin{algorithmic}[1]
\REQUIRE Evaluated variable space of Routine~(1) \STATE $\gamma
\Leftarrow \rhohat^T \rhat$ \STATE $\mathbf{u} \Leftarrow
(\rhohat-\gamma\rhat)/\|\rl{\mathbf{r}}\|$ \STATE $\mathbf{u}_1
\Leftarrow \mathbf{A}_1\mathbf{u}$ \STATE $\mathbf{u}_2 \Leftarrow
\mathbf{A}_2\mathbf{u}$ \STATE $\dt\sigma_1 \Leftarrow
-\sigma_1^3\mathbf{u}_1^T\mathbf{v}_1$ \STATE $\dt\sigma_2
\Leftarrow -\sigma_2^3\mathbf{u}_2^T\mathbf{v}_2$ \STATE
$\dt\rl{\mathbf{H}} \Leftarrow
-\frac{\dt\sigma_1}{\sigma_1^2}\mathbf{\Gamma}_1
-\frac{\dt\sigma_2}{\sigma_2^2}\mathbf{\Gamma}_2$ \STATE $\dt d_H
\Leftarrow 0$ \FOR{$i=1$ to $3$} \STATE $\mathbf{J} \Leftarrow
\rl{\mathbf{H}}^{(i)}$ \COMMENT{Defined in the corresponding
section} \STATE $\dt d_H \Leftarrow \dt d_H + \det\mathbf{J}$
\ENDFOR \STATE $\dt\rl{\eta} \Leftarrow
\rl{\eta}\frac{\dt\sigma_1+\dt\sigma_2}{2(\sigma_1+\sigma_2)}
-\frac{\rl{\eta}\dt d_H}{2 d_H} -\frac{2
\rl{\eta}}{\lambda}\big(\frac{d_{S1}\dt\sigma_1}{\sigma_1^3}+
\frac{d_{S1}\dt\sigma_1}{\sigma_1^3}\big)$ \STATE $\dt\rl{\chi}
\Leftarrow 4\mathbf{u}^T\mathbf{w}$ \STATE $\dt\rl{h} \Leftarrow
\gamma + \frac{1}{2}\rl{\sigma}^3\mathbf{u}^T\mathbf{s}$ \STATE
Evaluate $\dt U_{\alpha}$ using Eq.~(\ref{eq:dU})
\end{algorithmic}

\newpage

\newpage

\newpage

\section{Conclusion}
We have derived analytical expressions for the forces and torques
exerted on two molecules interacting via the \resq potential.
Moreover, efficient routines have been provided for molecular
dynamics simulations. A numerical investigation reveals that the
provided routines are 1.6 times faster than a two-point finite
difference approach. The evaluation of energy derivatives is the
most expensive element in a MD simulation. Using the provided
analytic derivatives, a MD simulation will run almost as fast as a
similar MC simulation (Eq.~\ref{MCMD}). This speedup leads to the
possibility of larger scale MD simulations of a wide range of
materials such as liquid crystals and certain organic molecules.

\section{Acknowledgement}
M. R. Ejtehadi would like to thank Institute for studies in
Theoretical Physics and Mathematics for partial supports.

\bibliography{re2deriv}

\begin{thebibliography}{10}
\expandafter\ifx\csname url\endcsname\relax
  \def\url#1{\texttt{#1}}\fi
\expandafter\ifx\csname urlprefix\endcsname\relax\def\urlprefix{URL }\fi

\bibitem{AllenTildesley}
M.~Allen, D.~Tildesley, Computer Simulation of Liquids, Oxford University
  Press, Oxford, 1989.

\bibitem{Frenkel}
D.~Frenkel, B.~Smit, Understanding Molecular Simulations, Academic Press, New
  York, 2002.

\bibitem{Leach}
A.~R. Leach, Molecular Modelling: Principles and Applications, Addison Wesley
  Longman Limited, 1996.

\bibitem{GB}
J.~G. Gay, B.~J. Berne, Modification of the overlap potential to mimic a linear
  site–site potential, J. Chem. Phys. 74 (1981) 3316--3319.

\bibitem{BFZ98}
R.~Berardi, C.~Fava, C.~Zannoni, A gay-berne potential for dissimilar biaxial
  particles, Chem. Phys. Lett. 297 (1998) 8--14.

\bibitem{Perram96}
J.~W. Perram, J.~Rasmussen, E.~Praestgaard, J.~L. Lebowitz, Ellipsoid contact
  potential: Theory and relation to overlap potentials, Phys. Rev. E 54 (1996)
  6565--6572.

\bibitem{Hunter}
R.~J. Hunter, Foundations of Colloid Science, Oxford University Press, USA,
  2001.

\bibitem{Babadi2006}
M.~Babadi, R.~Everaers, M.~R. Ejtehadi, Coarse-grained interaction potentials
  for anisotropic molecules, To be appeared on J. Chem. Phys.
  [cond-mat/0602308].

\bibitem{EE03}
R.~Everaers, M.~R. Ejtehadi, Interaction potentials for soft and hard
  ellipsoids, Phys. Rev. E 67 (2003) 041710.

\bibitem{Derjaguin}
B.~V. Derjaguin, Kolloid Z. 69 (1934) 155.

\bibitem{RE2CLib}
The parameterization routine (MATLAB/Octave) among with an efficient set of C
  subroutines for the evaluation of RE$^2$ interaction potential and its
  analytic derivatives are freely available at {\bf
  http://mehr.sharif.edu/~softmatter/RE-squared}.

\end{thebibliography}

\end{document}